\documentclass[sigconf,nonacm]{acmart}

\renewcommand\footnotetextcopyrightpermission[1]{}
\usepackage{amsmath}
\usepackage{graphicx}
\usepackage{xcolor}
\usepackage{enumerate}
\usepackage{booktabs}
\usepackage{multirow}
\usepackage{algorithm}
\usepackage{algpseudocode}
\usepackage{colortbl}
\usepackage{listings}
\usepackage{balance}

\definecolor{codegray}{rgb}{0.5,0.5,0.5}
\definecolor{codegreen}{rgb}{0,0.6,0}
\definecolor{codered}{rgb}{0.8,0,0}
\definecolor{codepurple}{rgb}{0.58,0,0.82}
\definecolor{backcolour}{rgb}{0.95,0.95,0.95}

\lstdefinestyle{codestyle}{
  basicstyle=\ttfamily\fontsize{7}{8.5}\selectfont,
  backgroundcolor=\color{backcolour},
  commentstyle=\color{codegreen},
  keywordstyle=\color{blue},
  numberstyle=\tiny\color{codegray},
  stringstyle=\color{codepurple},
  frame=lines,
  numbers=left,
  numbersep=5pt,
  breaklines=true,
  breakatwhitespace=true,
  tabsize=2,
  xleftmargin=10pt,
  framexleftmargin=8pt,
  columns=flexible,
  keepspaces=true,
  showstringspaces=false,
  captionpos=b,
  moredelim=[is][\color{codered}]{@DEL@}{@ENDDEL@},
  moredelim=[is][\color{codegreen}]{@ADD@}{@ENDADD@}
}
\usepackage{ifthen}

\newboolean{showcomments}
\setboolean{showcomments}{true}

\ifthenelse{\boolean{showcomments}}{%
  \newcommand{\mynote}[2]{%
    \fbox{\bfseries\sffamily\scriptsize #1}%
    {\small
      $\blacktriangleright$
      \textsf{\textcolor{red}{#2}}%
      $\blacktriangleleft$}%
  }%
}{%
  \newcommand{\mynote}[2]{}% empty definition when comments are off
}

\definecolor{bgcolor}{rgb}{0.95,0.95,0.95} 

\definecolor{white}{RGB}{255, 255, 255}         
\definecolor{lightgreen1}{RGB}{240, 255, 240}  
\definecolor{lightgreen2}{RGB}{200, 255, 200}  
\definecolor{lightgreen3}{RGB}{150, 255, 150} 
\definecolor{lightgreen4}{RGB}{100, 255, 100}  
\definecolor{lightgreen5}{RGB}{50, 200, 50}

\begin{document}

%% Title
\title{Diverse LLMs vs. Vulnerabilities: Who Detects and Fixes Them Better?}

%% Authors - use anonymous for review
\author{Arastoo Zibaeirad}
\affiliation{%
  \institution{University of North Carolina at Charlotte}
  \city{Charlotte}
  \state{NC}
  \country{USA}
}
\email{azibaeir@charlotte.edu}

\author{Marco Vieira}
\affiliation{%
  \institution{University of North Carolina at Charlotte}
  \city{Charlotte}
  \state{NC}
  \country{USA}
}
\email{marco.vieira@charlotte.edu}

%% Abstract and keywords
\begin{abstract}

Large Language Models (LLMs) are increasingly being studied for Software Vulnerability Detection (SVD) and Repair (SVR). Individual LLMs have demonstrated code understanding abilities, but they frequently struggle when identifying \textit{complex} vulnerabilities and generating fixes. 
This study presents \textbf{DVDR-LLM}, an ensemble framework that combines outputs from \textit{diverse} LLMs to determine whether aggregating multiple models reduces error rates. Our evaluation reveals that DVDR-LLM achieves 10-12\% higher detection accuracy compared to the average performance of individual models, with benefits increasing as code complexity grows. For multi-file vulnerabilities, the ensemble approach demonstrates significant improvements in recall (+18\%) and F1 score (+11.8\%) over individual models. However, the approach raises measurable trade-offs: reducing false positives in verification tasks while simultaneously increasing false negatives in detection tasks, requiring careful decision on the required level of agreement among the LLMs (threshold) for increased performance across different security contexts. \textbf{Artifact:} \url{https://github.com/Erroristotle/DVDR_LLM}
\end{abstract}

%% CCS Concepts


%% Keywords

%% Generate title and metadata
\maketitle

\section{Introduction}
Software Vulnerability Detection (SVD) and Software Vulnerability Repair (SVR) are increasingly challenging as modern software grows in complexity and interconnectedness. Identifying and addressing vulnerabilities require tools that can comprehensively understand code semantics, manage the complexities of evolving structures, and recognize the intricate connections and dependencies among code components \cite{zhou2024large}.

While traditional Static Analysis Tools (SATs) are useful in early software development stages, they are limited in handling complex, dynamic environments. Unlike dynamic analysis tools, which execute code to identify runtime vulnerabilities and excel in detecting runtime-related issues, SATs only have a static view of the code. They analyze abstract information, such as control flow, data flow, and function calls, but often fail to capture the nuances of context-sensitive vulnerabilities. This limitation results in a high rate of false positives and false negatives, particularly for more intricate vulnerabilities \cite{pearce2022asleep}.

Recent research has explored LLMs as tools for SVD and SVR, leveraging their capacity to identify patterns within vast code and vulnerability datasets \cite{mao2024towards,liu2024source}. These models provide a complementary analysis layer, supporting the detection of risky code constructs and suggesting potential repair strategies. However, individual LLMs face significant challenges in reliably detecting and repairing vulnerabilities, particularly in complex, context-sensitive cases \cite{zibaeirad2024vulnllmeval}. They often struggle to accurately distinguish vulnerable from non-vulnerable code, leading to simplistic or overly intricate patch suggestions that may fail to resolve underlying issues. %These limitations can leave vulnerabilities unresolved or introduce new issues. 

While previous studies have primarily evaluated individual LLMs for security tasks or explored ensemble methods in general NLP tasks, our work is the first comprehensive empirical evaluation of how model diversity specifically impacts vulnerability detection and repair. In fact, unlike prior research that typically focuses on improving individual model performance, we examine the trade-offs in aggregating outputs from diverse model families and parameter sizes. This novel approach reveals critical insights about precision-recall balancing across different vulnerability types and abstraction levels that cannot be observed when studying models in isolation.

To explore the practical effectiveness of this multi-model approach, we designed DVDR-LLM, a framework that leverages the \underline{\textbf{D}}iversity of multiple LLMs and their agreement via a consensus threshold for improved \underline{\textbf{V}}ulnerability \underline{\textbf{D}}etection and \underline{\textbf{R}}epair. Building on the principles of ensemble learning, which combine multiple models to reduce errors and improve generalization\cite{dietterich2002ensemble}, DVDR-LLM enhances the understanding of complex vulnerability contexts and identifies practical trade-offs. Our contribution is not simply proposing another detection system but rather providing a systematic analysis of how diversity and consensus thresholds affect security-critical tasks across different code complexity levels, with clear evidence that ensemble benefits increase with code complexity.

\textbf{Research Gaps.} From a literature review, we identify three critical gaps: 
(1) existing approaches rely heavily on individual LLMs despite their documented inconsistency in security-critical tasks;
(2) the impact of model diversity on security task performance remains unexplored, particularly as code complexity increases; and
(3) there is limited understanding of how consensus thresholds affect the precision-recall trade-off in vulnerability detection and repair scenarios.

\textbf{Research Questions.} To address these gaps, we explore the following research questions:

\begin{itemize}
    \item \textbf{RQ1:} How does aggregating multiple LLMs impact SVD performance compared to individual models? (Gap 1)
    \item \textbf{RQ2:} What is the best aggregation strategy threshold for balancing precision and recall in an ensemble? (Gap 3)
    \item \textbf{RQ3:} How does the ensemble-based system perform across different abstraction levels? (Gap 2)
    \item \textbf{RQ4:} How does weighted aggregation affect the quality of code repairs? (Gaps 1 and 2)
\end{itemize}

The remainder of this paper is structured as follows: 
\textbf{Section \ref{section:approach}} introduces DVDR-LLM, outlining its design and methodology. \textbf{Section \ref{section:setup}} describes the experimental setup, while \textbf{Section \ref{section:results}} presents the findings and answers to the RQs. \textbf{Section \ref{section:discussion}} discusses the limitations of our work and future research directions. 
\textbf{Section \ref{section:related-works}} discusses related studies, and finally, \textbf{Section \ref{section:conclusion}} concludes the paper.

\begin{figure*}[htbp]
    \centerline{\includegraphics[width=0.85\textwidth]{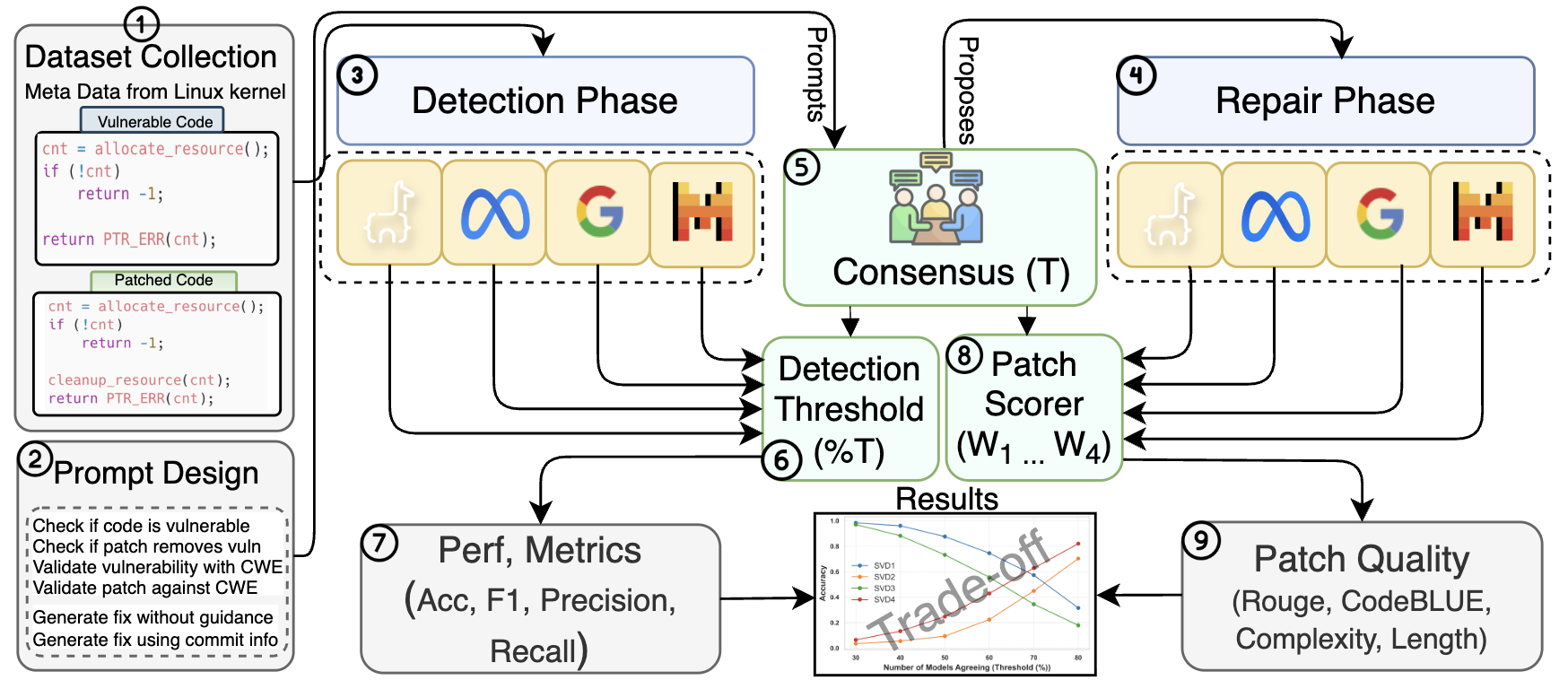}}
    \caption{DVDR-LLM Framework. 
    \textnormal{\textcircled{\scriptsize 1}} Real-world vulnerabilities are collected and paired as vulnerable/patched code blocks. 
    \textnormal{\textcircled{\scriptsize 2}} Prompt templates guide LLMs for detection and repair tasks. 
    \textnormal{\textcircled{\scriptsize 3}} Multiple LLMs independently perform vulnerability detection (SVD). 
    \textnormal{\textcircled{\scriptsize 4}} The same models propose code repairs (SVR). 
    \textnormal{\textcircled{\scriptsize 5}} Outputs are aggregated via a consensus mechanism. 
    \textnormal{\textcircled{\scriptsize 6}} A threshold $T$ determines ensemble agreement for detection. 
    \textnormal{\textcircled{\scriptsize 7}} Performance is evaluated using accuracy, precision, recall, and F1 score. 
    \textnormal{\textcircled{\scriptsize 8}} Patch candidates are ranked using a weighted scorer. 
    \textnormal{\textcircled{\scriptsize 9}} Final results reveal trade-offs across models, tasks, and thresholds.}
    \label{figure:framework}
\end{figure*}

\section{DVDR-LLM}
\label{section:approach}
DVDR-LLM employs an ensemble-based approach for SVD and SVR. As shown in Figure \ref{figure:framework}, multiple LLMs independently assess code with a vulnerability flagged if a majority of models agree. For repair, each model proposes fixes and the framework selects the optimal patch using a weighted scoring system.
Model diversity generally provides three key benefits:
\begin{enumerate}
    \item \textbf{Complementary Detection}: Different model families excel at detecting different vulnerability types \cite{xu2022systematic}.
    \item \textbf{Size Optimization}: Larger models perform better at verification, while smaller models have higher detection sensitivity \cite{xu2022systematic}.
    \item \textbf{Context Handling}: Models with varying context windows complement each other on complex vulnerabilities \cite{xu2022systematic,pearce2025asleep}.
\end{enumerate}

\subsection{Detection Phase}
A consensus threshold \( T \) balances precision and recall. Let \( M_1, M_2, \dots, M_n \) be the outputs of \( n \) LLMs, where \( M_i \in \{0,1\} \) indicates vulnerability detection. Ensemble decision is:

\begin{equation}
f(M_1, M_2, \dots, M_n) =
\begin{cases}
1 & \text{if } \sum_{i=1}^{n} M_i \geq T, \\
0 & \text{otherwise}.
\end{cases}
\end{equation}

Different thresholds can be applied based on security requirements: higher for critical systems needing near-zero false positives, lower where recall is prioritized.

\subsection{Repair Phase}
\label{subsec:repair}

Each LLM proposes a fix, and the framework selects the patch with the highest score \( S \). The score is a weighted combination of metrics on key aspects of patch quality:

\begin{equation}
\small
S = w_1 \cdot \text{ROUGE} + w_2 \cdot \text{CodeBLEU} + w_3 \cdot \text{Complexity} + w_4 \cdot \text{Length},
\end{equation}

Weights \( (w_1 = w_2 = 0.2, w_3 = w_4 = 0.3) \) were determined empirically on real-world patches. Surface similarity metrics (ROUGE and CodeBLEU) receive lower weights (0.2) as they do not fully represent patch effectiveness, while quality metrics (Complexity and Length) receive higher weights (0.3) to prioritize maintainable patches that more closely match the complexity profile of human patches, which our analysis showed is critical for security-sensitive code.
We emphasize that Cyclomatic Complexity serves as a control-flow metric that relates to maintainability rather than definitive code quality. Both excessive and insufficient complexity can be problematic for security patches. Detailed explanations of the metrics are provided in Section~\ref{subsection:metrics}.

\section{Experimental Setup}
\label{section:setup}

This section details our experimental setup for evaluating LLM diversity in vulnerability detection and repair. We begin by describing the dataset, highlighting its structure and relevance to SVD and SVR. Section~\ref{section:models} covers the LLMs used, while Section~\ref{subsection:metrics} outlines the performance metrics for accuracy, robustness, and relevance. Section~\ref{subsection:prompts} concludes with the prompt templates used to standardize inputs.

\subsection{Dataset}
The dataset utilized in this study is sourced from VulnLLMEval \cite{zibaeirad2024vulnllmeval}, which focuses on real-world Linux kernel vulnerabilities to evaluate LLMs in SVD and SVR tasks. VulnLLMEval systematically gathers commit hashes from authoritative vulnerability advisory resources, such as the NVD (National Vulnerability Database), that explicitly link commits to CVE and CWE identifiers. These links are verified against security databases, ensuring that each commit indeed corresponds to a security fix.

To establish ground truth, for each CVE-linked commit (i.e., the fix), the dataset includes both the patched and the immediately preceding version of the affected code. The version before the patch is labeled as \textit{vulnerable}, and the patched version is labeled as \textit{non-vulnerable}, under the common assumption in vulnerability research that the fix directly addresses the vulnerability described by the CVE. 
%\mv{mention that this is a common approach in this type of study; include a citation}. 
This pairing is done at fine-grained levels, including file-, function-, and non-functional code blocks, allowing for structured comparison and model evaluation across varying abstraction levels.

To support fix correction in few-shot prompting scenarios, the dataset includes commit descriptions from verified security fixes. It is important to note that these are not \textit{potential} fixes but rather KNOWN and VERIFIED fixes that have been explicitly linked to specific CVEs by security authorities and confirmed to address the vulnerabilities in question. These commit descriptions contain domain expert knowledge about the vulnerability and its resolution, offering valuable guidance to the LLMs for generating appropriate patches.

We acknowledge that such heuristics may introduce some degree of noise; for example, commits may occasionally include refactoring, comments, or unrelated changes alongside the vulnerability fix. However, the reliance on explicitly documented CVE-associated commits substantially reduces the likelihood of arbitrary or misclassified labeling. Furthermore, by restricting the dataset to cases with well-annotated CVE and CWE context, VulnLLMEval provides a focused and reliable framework for studying SVD and SVR in real-world scenarios.

\subsection{Models}
\label{section:models}
DVDR-LLM utilizes a diverse set of pre-trained LLMs, including CodeLlama \cite{roziere2023code}, Gemma2 \cite{team2024gemma}, Llama3, Llama3.1 \cite{dubey2024llama}, and Mistral \cite{jiang2023mistral}, as summarized in Table \ref{table:llms}. These models vary in size, context window length (8,192 to 131,072 tokens), and instruction-tuning capabilities, allowing for comprehensive evaluation across different detection and repair scenarios.

\begin{table}[b]
    \centering
    \captionsetup{skip=2pt}
    \caption{LLM Parameters and Specifications}
    \label{table:llms}
    \scriptsize
    \setlength{\tabcolsep}{4pt}
    \begin{tabular}{llr}
    \toprule
    \textbf{Model Class} & \textbf{Model Version} & \textbf{Context (Tokens)} \\ 
    \midrule
    \multirow{2}{*}{\textbf{CodeLlama}} 
        & 7B-instruct & 16,384 \\ 
        & 34B-instruct & 16,384 \\
    \textbf{Llama3} 
        & 70B-instruct & 8,192 \\
        & 8B-instruct & 8,192 \\ 
    \textbf{Llama3.1} 
        & 70B & 131,072 \\
        & 8B & 131,072 \\ 
    \textbf{Gemma2} 
        & 27B & 8,192 \\ 
        & 9B & 8,192 \\ 
    \textbf{Mistral} 
        & 7B-instruct & 32,768 \\ 
        & Mixtral-8*7B & 32,768 \\ 
    \bottomrule
    \end{tabular}
\end{table}

All models were accessed via their official API endpoints with default temperature settings (0.7) and top-p of 0.9, controlling the randomness and diversity of the outputs. We did not perform any additional fine-tuning to maintain a fair comparison of the models' base capabilities. The models represent a diverse range of architectures and training approaches.

\begin{table*}[t]
    \centering
    \captionsetup{skip=4pt}
    \caption{Prompt Templates for Vulnerability Detection and Repair Tasks}
    \resizebox{\textwidth}{!}{%
        \begin{tabular}{c|p{0.8cm}p{2.5cm}p{2cm}p{9.5cm}} 
        \toprule
        \textbf{Tasks} & \textbf{ID} & \textbf{Task Type} & \textbf{Inputs} & \textbf{Description} \\ 
        \midrule
        
        \multirow{4}{*}{\rotatebox[origin=c]{90}{\textbf{SVD}}} 
        & SVD1 & Vulnerability Check & V & Determines if the unpatched code block contains vulnerabilities \\ 
        \cmidrule{2-5}
        & SVD2 & Patch Validity Check & P & Evaluates whether the patched code block successfully removes the vulnerabilities \\ 
        \cmidrule{2-5}
        & SVD3 & Vulnerability Check & V, CVE, CWE & Checks if the unpatched code aligns with the given CVE/CWE classification \\ 
        \cmidrule{2-5}
        & SVD4 & Patch Validity Check & P, CVE, CWE & Verifies whether the patched code properly addresses the corresponding CVE/CWE \\ 
        \midrule
        
        \multirow{2}{*}{\rotatebox[origin=c]{90}{\textbf{SVR}}} 
        & SVR1 & Fix Suggestion (Z) & V & Generates a potential fix for the vulnerability without external guidance \\ 
        \cmidrule{2-5}
        & SVR2 & Fix Suggestion (F) & V, D, CVE, CWE & Creates a patch using commit descriptions and CWE/CVE context as guidance \\
        \bottomrule
        \end{tabular}
    }
    \vspace{-2mm}
    \begin{center}
    \small{Z = Zero-shot; F = Few-shot; V = Vulnerable code; P = Patched code; D = Commit description}
    \end{center}
    \label{table:prompt_templates}
\end{table*}

\subsection{Metrics}
\label{subsection:metrics}
We employ common metrics to assess accuracy, relevance, and robustness in SVD and SVP tasks:

%\subsubsection{SVD Metrics}

\noindent\textit{1. SVD Metrics}
\begin{enumerate}
    \item \textit{F1 Score, Recall, Accuracy, Precision}: Provide foundational assessment of detection accuracy, balancing true/false positives and negatives.
\end{enumerate}

%\noindent\subsubsection{SVP Metrics}
\noindent\textit{2. SVP Metrics}
\begin{enumerate}

    \item \textit{ROUGE Score} \cite{lin2004rouge}: Evaluates similarity between generated and reference patches by n-gram overlap, with ROUGE-L capturing the longest common subsequence. Useful for measuring structural alignment, but it may overlook the logical accuracy required in code fixes.
    
    \item \textit{CodeBLEU} \cite{ren2020codebleu}: Designed for code evaluation, it extends traditional BLEU \cite{papineni2002bleu} by assessing syntactic and semantic accuracy, crucial for meaningful patch assessment. By evaluating surface similarity along with structural correctness, addresses the limitations of simpler text metrics to capture syntax, semantics, and logical correctness.

    \item \textit{Cyclomatic Complexity} \cite{mccabe1976complexity}: Counts independent paths in code, reflecting logical complexity. For SVP, it provides insights into maintainability by comparing complexity pre- and post-patch, highlighting if generated patches introduce unnecessary complexity or simplify the code.

    \item \textit{Code Length}: Longer code often indicates more thorough fixes for complex vulnerabilities. As a straightforward measure, Code Length provides insight into patch scope, though it may not fully capture maintainability if excessive length adds unnecessary complexity.

\end{enumerate}

\subsection{Prompt Templates}
\label{subsection:prompts}
We systematically designed prompt templates for zero-shot scenarios in SVD and both zero-shot and few-shot scenarios in SVR. The design process was informed by prior research and iterative testing to ensure clarity and alignment with task objectives \cite{ullah2024llms,zibaeirad2024vulnllmeval}. Note that, we excluded few-shot prompts from SVD tasks to prevent biasing models toward specific vulnerability patterns. For SVR tasks, we included few-shot examples as they guide repair approaches without compromising evaluation validity. Table \ref{table:prompt_templates} summarizes the types, inputs, and purposes of each prompt.

Our approach evaluates the LLMs' ability to detect vulnerabilities and verify patches using Yes/No question-based prompts. Prompts SVD1 and SVD2 assess whether unpatched and patched code blocks, respectively, contain vulnerabilities. Prompts SVD3 and SVD4 extend this by testing the models' ability to link vulnerabilities and patches to specific CVE or CWE identifiers, ensuring alignment with ground truth and real-world security standards.

For the patching tasks, we assess the LLMs' ability to generate effective fixes for vulnerabilities in code blocks through two settings: zero-shot (SVR1) and few-shot (SVR2). In the first, the model generates a patch independently, relying solely on its training data and knowledge. In the second, a commit description is provided to guide its patch generation.

\section{Experimental Results}
\label{section:results}

\subsection{RQ1: DVDR-LLM's Impact on SVD Performance}
Aggregating LLMs combines their outputs to enhance performance in SVD. A 70\% consensus threshold was used for our initial evaluation. As Table \ref{tab:model_comparison_delta} shows, ensemble effectiveness varies across tasks, with SVD1 and SVD3 showing more false negatives, while SVD2 and SVD4 fewer false positives.

For vulnerability detection tasks (SVD1 and SVD3), most individual models outperform the ensemble. For SVD1, 7 of 10 models show better performance, with Llama3-8b showing remarkably higher detection rates (+59.8\%), followed by Llama3.1-8b (+52.5\%) and Gemma2-9b (+46.4\%). The pattern is even more pronounced for SVD3, where 9 of 10 models outperform the ensemble, with Llama3.1-8b (+130.8\%), Llama3-8b (+127.1\%), and Gemma2-9b (+111.2\%) showing more than double the detection capability.

The data reveals a consistent inverse relationship: models with superior vulnerability detection typically show poor patch verification capability. For example, Llama3.1-8b excels at detection (SVD1: +52.5\%, SVD3: +130.8\%) but performs poorly at verification (SVD2: -74.3\%, SVD4: -69.7\%). Conversely, Llama3-70b struggles with detection (SVD1: -50.8\%, SVD3: -19.6\%) but excels at verification (SVD2: +54.4\%, SVD4: +13.5\%). The ensemble effectively balances these extremes, prioritizing false positive reduction in verification tasks at the cost of increased false negatives in detection tasks.

\begin{table}[b]
    \centering
    \renewcommand{\arraystretch}{1.0}
    \setlength{\tabcolsep}{2pt}
    \caption{Comparison of Model Performance on SVD Tasks with Delta Percentages from \textbf{DVDR-LLM (our proposed ensemble)} Baseline (Threshold = 70\%).}
    \label{tab:model_comparison_delta}
    \large
    \resizebox{\columnwidth}{!}{%
    \begin{tabular}{l|cc|cc|cc|cc}
    \toprule
    \multirow{2}{*}{\textbf{Model}} & \multicolumn{2}{c|}{\textbf{SVD1}} & \multicolumn{2}{c|}{\textbf{SVD2}} & \multicolumn{2}{c|}{\textbf{SVD3}} & \multicolumn{2}{c}{\textbf{SVD4}} \\
    \cmidrule(lr){2-3} \cmidrule(lr){4-5} \cmidrule(lr){6-7} \cmidrule(lr){8-9}
     & \textbf{Correct} & \textbf{$\mathbf{\Delta}$ (\%)} & \textbf{Correct} & \textbf{$\mathbf{\Delta}$ (\%)} & \textbf{Correct} & \textbf{$\mathbf{\Delta}$ (\%)} & \textbf{Correct} & \textbf{$\mathbf{\Delta}$ (\%)} \\
    \midrule
    CodeLlama-7b     & 209 & \cellcolor{green!15}+16.8\%  & 89  & \cellcolor{red!15}-34.6\%  & 200 & \cellcolor{green!25}+86.9\%  & 101 & \cellcolor{red!25}-45.4\% \\
    CodeLlama-34b    & 185 & \cellcolor{green!10}+3.4\%   & 119 & \cellcolor{red!10}-12.5\%  & 157 & \cellcolor{green!20}+46.7\%  & 147 & \cellcolor{red!15}-20.5\% \\
    Llama3-8b        & 286 & \cellcolor{green!25}+59.8\%  & 17  & \cellcolor{red!30}-87.5\%  & 243 & \cellcolor{green!30}+127.1\% & 49  & \cellcolor{red!30}-73.5\% \\
    Llama3-70b       & 88  & \cellcolor{red!25}-50.8\%  & 210 & \cellcolor{green!25}+54.4\%  & 86  & \cellcolor{red!15}-19.6\%  & 210 & \cellcolor{green!15}+13.5\% \\
    Llama3.1-8b      & 273 & \cellcolor{green!25}+52.5\%  & 35  & \cellcolor{red!30}-74.3\%  & 247 & \cellcolor{green!30}+130.8\% & 56  & \cellcolor{red!30}-69.7\% \\
    Llama3.1-70b     & 211 & \cellcolor{green!15}+17.9\%  & 100 & \cellcolor{red!15}-26.5\%  & 205 & \cellcolor{green!25}+91.6\%  & 95  & \cellcolor{red!25}-48.6\% \\
    Mistral-7b       & 158 & \cellcolor{red!10}-11.7\%  & 145 & \cellcolor{green!10}+6.6\%   & 121 & \cellcolor{green!15}+13.1\%  & 184 & \cellcolor{red!5}-0.5\%  \\
    Mixtral-8*7b     & 224 & \cellcolor{green!15}+25.1\%  & 77  & \cellcolor{red!20}-43.4\%  & 110 & \cellcolor{green!10}+2.8\%   & 194 & \cellcolor{green!10}+4.9\%  \\
    Gemma2-9b        & 262 & \cellcolor{green!20}+46.4\%  & 53  & \cellcolor{red!25}-61.0\%  & 226 & \cellcolor{green!30}+111.2\% & 75  & \cellcolor{red!25}-59.5\% \\
    Gemma2-27b       & 146 & \cellcolor{red!15}-18.4\%  & 174 & \cellcolor{green!15}+27.9\%  & 183 & \cellcolor{green!25}+71.0\%  & 121 & \cellcolor{red!20}-34.6\% \\
    \midrule
    \rowcolor{gray!15} DVDR-LLM & 179 & --- & 136 & --- & 107 & --- & 185 & --- \\
    \bottomrule
    \end{tabular}
    }\vspace{0.5em}
    \scriptsize{In each column, \textbf{Correct} shows correctly classified samples, \textbf{$\mathbf{\Delta}$ (\%)} shows percentage difference from DVDR-LLM. Green/red cells indicate better/worse performance.}
\end{table}

DVDR-LLM has its most significant impact on patch verification tasks (SVD2 and SVD4), where it substantially reduces false positives compared to many individual models. Looking at the delta percentages, we observe that for SVD2, smaller models like Llama3-8b and Llama3.1-8b dramatically underperform the ensemble baseline (by -87.5\% and -74.3\% respectively), with Gemma2-9b also showing poor performance (-61.0\%). Similarly, for SVD4, these same models show substantial underperformance: Llama3-8b (-73.5\%), Llama3.1-8b (-69.7\%), and Gemma2-9b (-59.5\%). Overall, the ensemble outperforms 7 of the 10  models on both SVD2 and SVD4.

\begin{figure}[ht]
\centering
\begin{lstlisting}[style=codestyle]
// File path: fs/ext4/ext4.h
static inline loff_t ext4_isize(struct ext4_inode *raw_inode)
{
@DEL@-   return ((loff_t)le32_to_cpu(raw_inode->i_size_high) << 32) |@ENDDEL@
@DEL@-          le32_to_cpu(raw_inode->i_size_lo);@ENDDEL@
@ADD@+   if (S_ISREG(le16_to_cpu(raw_inode->i_mode)))@ENDADD@
@ADD@+       return ((loff_t)le32_to_cpu(raw_inode->i_size_high) << 32) |@ENDADD@
@ADD@+              le32_to_cpu(raw_inode->i_size_lo);@ENDADD@
@ADD@+   else@ENDADD@
@ADD@+       return (loff_t) le32_to_cpu(raw_inode->i_size_lo);@ENDADD@
}

// File path: fs/ext4/inode.c
static int ext4_block_to_path(struct inode *inode,
             ext4_lblk_t i_block,
             ext4_lblk_t offsets[4], int *boundary)
 {
 
 ...
 
} else {
@DEL@-   ext4_warning(inode->i_sb, "ext4_block_to_path",@ENDDEL@
@DEL@-                "block %lu > max",@ENDDEL@
@DEL@-                i_block + direct_blocks + indirect_blocks + double_blocks);@ENDDEL@
@ADD@+   ext4_warning(inode->i_sb, "ext4_block_to_path",@ENDADD@
@ADD@+                "block %lu > max in inode %lu",@ENDADD@
@ADD@+                i_block + direct_blocks + indirect_blocks + double_blocks,@ENDADD@
@ADD@+                inode->i_ino);@ENDADD@
}
\end{lstlisting}
\tiny
\caption{Patch for \textbf{CVE-2009-0747} (\textbf{CWE-399}): Subtle input validation vulnerability where using high bits of inode size for all file types created a denial-of-service opportunity. The fix restricts high-bit usage to regular files only, preventing excessive CPU consumption when processing malformed directories. %This example demonstrates why ensemble models sometimes struggle with subtle semantic vulnerabilities requiring domain-specific knowledge.
}
\label{fig:cve2009_patch}
\end{figure}

Figure \ref{fig:cve2009_patch} shows an example of the precision-recall trade-offs identified in RQ1. This subtle input validation vulnerability (CVE-2009-0747) achieved zero consensus among our models when using the 60\% threshold, meaning at least 6 models would need to agree for a positive detection. Examining the individual model performance reveals that only 2 models (CodeLlama-34b and Llama3-8b) correctly identified the vulnerable code as vulnerable (SVD1), while only 2 models (Mistral-7b and CodeLlama-7b) correctly identified the patched code as non-vulnerable (SVD2). For SVD3 and SVD4, only 4 and 5 models respectively made correct assessments. This perfectly demonstrates our quantitative findings from Table \ref{tab:model_comparison_delta}: the ensemble's threshold requirement eliminates these scattered correct detections from individual models, increasing false negatives in SVD1/SVD3 while reducing false positives in SVD2/SVD4. Notably, the inverse relationship pattern is visible even in this single example: Llama3-8b correctly performs SVD1 but fails on SVD2, while Mistral-7b succeeds on SVD2 but fails on SVD1. This highlights why ensemble approaches sometimes struggle with domain-specific vulnerabilities requiring specialized filesystem knowledge; precisely the cases where model diversity cannot overcome knowledge gaps shared across all models.

\subsection{RQ2: Threshold for Optimal Detection Accuracy}
\begin{figure}[b]
\centerline{\includegraphics[width=0.45\textwidth]{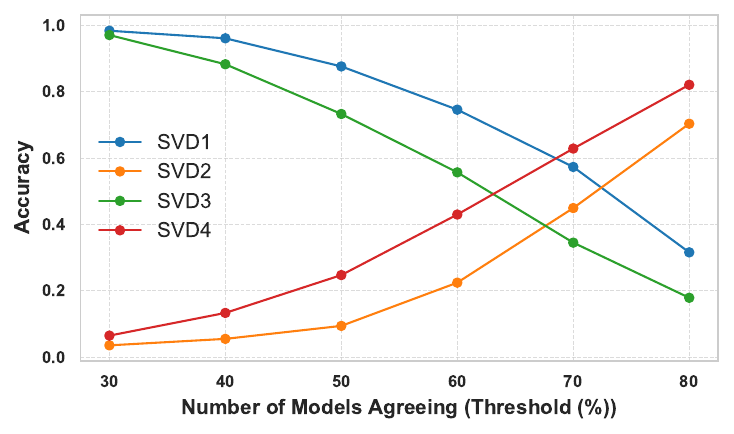}}
\caption{Accuracy of SVD Tasks at Varying Levels of Consensus}
\label{figure:RQ2}
\end{figure}

This research question examines how varying the threshold for model consensus impacts detection accuracy and precision-recall trade-offs. Threshold levels were evaluated from 30\% to 80\% in 10\% increments.
Figure \ref{figure:RQ2} shows that higher thresholds generally improve performance in patch verification tasks (SVD2 and SVD4) but cause slight accuracy declines in vulnerability detection (SVD1 and SVD3). 

These results reveal critical trade-offs when setting thresholds for ensemble consensus. While higher thresholds enhance precision and minimize omissions in patched code scenarios, they can hinder flexibility and decrease sensitivity in broader detection contexts. Tailoring thresholds to specific detection objectives rather than applying uniform criteria is essential to balance robustness and adaptability across diverse tasks.

While consensus thresholds below 50\% generally do not represent practical majority agreement, we present results across the entire threshold spectrum primarily to demonstrate accuracy trends. In practice, LLMs tend to label code as vulnerable more readily (tasks SVD1 and SVD3), whereas confirming that vulnerabilities have been properly patched (tasks SVD2 and SVD4) proves inherently more challenging. 

Based on the  findings above, we selected the 60\% threshold for all subsequent evaluations. This choice ensures an optimal balance across the diverse SVD tasks, tailoring the ensemble to meet both precision and recall requirements in practical applications. 

\subsection{RQ3: Performance across abstraction levels}
We analyzed metrics at three abstraction levels: Level 1 (single file, single function), Level 2 (single file, multiple functions), and Level 3 (multiple files, multiple functions). Our ensemble approach with a 60\% threshold consistently outperforms individual models as code complexity increases.

At Level 1, individual models performed comparably to the ensemble. However, at Level 3, the ensemble demonstrated significant improvements in recall (+18.05\%) and F1 score. The ensemble maintained stable accuracy (60.19\%-60.00\%) across all levels, consistently 10-12\% higher than average models. Also, while individual top performers achieved higher recall values (75.18\%-84.07\%), they sacrificed precision and overall accuracy. The ensemble prioritized balanced performance, achieving the highest F1 score (67.24\%) at Level 3. This demonstrates that model diversity benefits increase with abstraction level, becoming crucial for complex, multi-file vulnerabilities.

\begin{table*}[htbp]
    \centering
    \caption{Comparison of Average Metrics, Best Individual Models, and Ensemble Performance per Abstraction Level (Level 1: 1 file, 1 function; Level 2: 1 file, multiple functions; Level 3: multiple files, multiple functions). Bold values indicate best performance, underlined values indicate second-best. Values in parentheses represent improvements over average metrics. Results aggregated over SVD3 and SVD4 tasks with CWE guidance.}
    \resizebox{\textwidth}{!}{%
    \begin{tabular}{l|cccc|cccc|cccc}
        \toprule
        \multirow{2}{*}{\textbf{Abstraction}} & \multicolumn{4}{c|}{\textbf{Average Metrics (Mean ± Std)}} & \multicolumn{4}{c|}{\textbf{Best Individual Model}} & \multicolumn{4}{c}{\textbf{Ensemble Metrics (Threshold 60\%)}} \\
        \cmidrule(lr){2-5} \cmidrule(lr){6-9} \cmidrule(lr){10-13}
        \textbf{Level} & \textbf{Precision} & \textbf{Recall} & \textbf{Accuracy} & \textbf{F1 Score} & \textbf{Precision} & \textbf{Recall} & \textbf{Accuracy} & \textbf{F1 Score} & \textbf{Precision} & \textbf{Recall} & \textbf{Accuracy} & \textbf{F1 Score} \\
        \midrule
        Level 1 & 48.17±1.66 & 53.31±22.30 & 48.28±1.54 & 48.44±12.64 & \underline{50.24} & \textbf{75.18} & \underline{50.35} & \underline{60.00} & \textbf{60.58(+12.41)} & \underline{58.33(+5.02)} & \textbf{60.19(+11.91)} & \textbf{59.43(+10.99)} \\
        Level 2 & 49.99±4.98 & 57.41±20.57 & 49.99±4.26 & 51.98±10.43 & \underline{57.14} & \textbf{86.79} & \underline{57.00} & \underline{65.22} & \textbf{61.22(+11.23)} & \underline{65.22(+7.81)} & \textbf{61.96(+11.97)} & \textbf{63.16(+11.18)} \\
        Level 3 & 50.21±1.67 & 64.06±15.29 & 49.88±2.10 & 55.44±5.60 & \underline{52.50} & \textbf{84.07} & \underline{52.21} & \underline{62.50} & \textbf{56.93(+6.72)} & \underline{82.11(+18.05)} & \textbf{60.00(+10.12)} & \textbf{67.24(+11.80)} \\
        \bottomrule
    \end{tabular}
    }
    \label{tab:abstraction-level}
\end{table*}

Table \ref{tab:abstraction-level} reveals clear performance patterns across abstraction levels. The ensemble consistently outperforms average individual model metrics with precision improvements of 12.41\%, 11.23\%, and 6.72\% for Levels 1-3. While precision advantage decreases with higher abstraction, recall shows the opposite trend, improving from +5.02\% at Level 1 to +18.05\% at Level 3. The ensemble maintains stable accuracy (60.19\%, 61.96\%, 60.00\%) across all levels, consistently 10-12\% higher than average models.

Individual top performers achieve higher recall values (75.18\%-84.07\%) but sacrifice precision and overall accuracy. The ensemble prioritized balanced performance, achieving the highest F1 score (67.24\%) at Level 3, even surpassing the best individual model (62.50\%). This demonstrates that model diversity benefits increase with abstraction level. While the performance gap is modest at Level 1, by Level 3 the ensemble's multiple perspectives become crucial for complex, multi-file vulnerabilities, providing more reliable assessment where single models often fail.

\subsection{RQ4: Impact of Weighted Aggregation on Patch Quality}

\subsubsection{Similarity between generated and patched code}

\begin{figure}[bp]
\centerline{\includegraphics[width=0.5\textwidth]{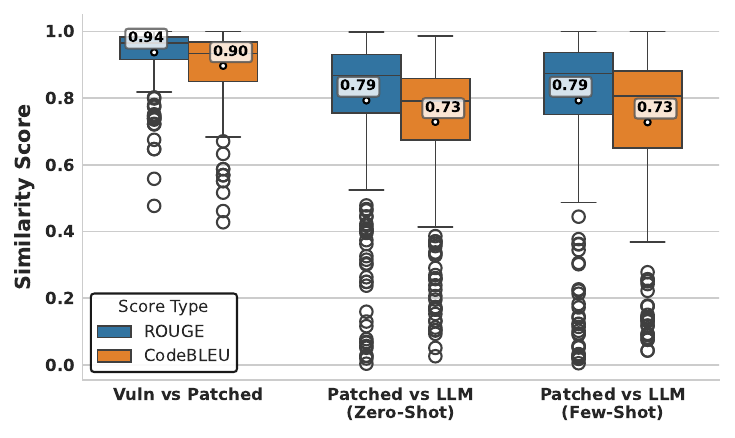}}
\caption{ROUGE and CodeBLEU similarity score distributions for vulnerable vs. patched code, and patched vs. LLM-generated patches (Zero-Shot and Few-Shot).}
\label{figure:RQ4-similarity}
\end{figure}

We evaluated patch quality using ROUGE and CodeBLEU metrics across three comparison scenarios (see Figure \ref{figure:RQ4-similarity}):

\begin{itemize}
    % \item \textit{Vulnerable Code vs. Patched Code:} Baseline to measure the structural and semantic differences between the original vulnerable code and the human-created patch. The similarity scores provide a baseline for understanding how much transformation the original code undergoes during patching. ROUGE scores here indicate significant structural changes, as expected, while CodeBLEU captures differences in syntax and logic, which are often more pronounced.
    \item \textit{Vulnerable Code vs. Patched Code:} Baseline for assessing both structural and semantic changes introduced by human-generated patches. Similarity scores reflect the degree of transformation the original code undergoes during patching. As expected, ROUGE scores indicate substantial structural modifications, while CodeBLEU highlights more nuanced changes in syntax and logic.
    
    \item \textit{Patched Code vs. LLM-Generated Patches (Zero-Shot):} The similarity scores for zero-shot patches indicate a moderate structural resemblance to human-created patches. ROUGE scores reflect reasonable structural overlap, while lower CodeBLEU scores reveal challenges in replicating the precise syntax and logical flow. This suggests that although zero-shot LLM-generated patches can approximate human structural modifications, they often fall short in capturing the nuanced semantics and syntactic accuracy of expert-written patches.
    
    \item \textit{Patched Code vs. LLM-Generated Patches (Few-Shot):} Similarity between the human patch extracted from the repository and those generated in the few-shot setting, where the model is guided by commit descriptions of the patches to fix the code. Few-shot patches consistently outperform zero-shot patches in both ROUGE and CodeBLEU scores. The improvement is more pronounced in CodeBLEU, indicating that providing contextual examples helps the model better capture the logical structure and syntactic nuances of the patched code. This demonstrates that few-shot learning enhances the semantic and syntactic quality of generated patches, bringing them more closer to human-created ones.
\end{itemize}

CodeBLEU scores are generally lower than ROUGE scores across all scenarios because CodeBLEU evaluates deeper aspects like syntax, semantics, and logic rather than just surface-level structural similarities, highlighting where LLM-generated patches deviate from intended logical corrections.

\subsubsection{Complexity Analysis}
\begin{figure}[tbp]
\centerline{\includegraphics[width=0.5\textwidth]{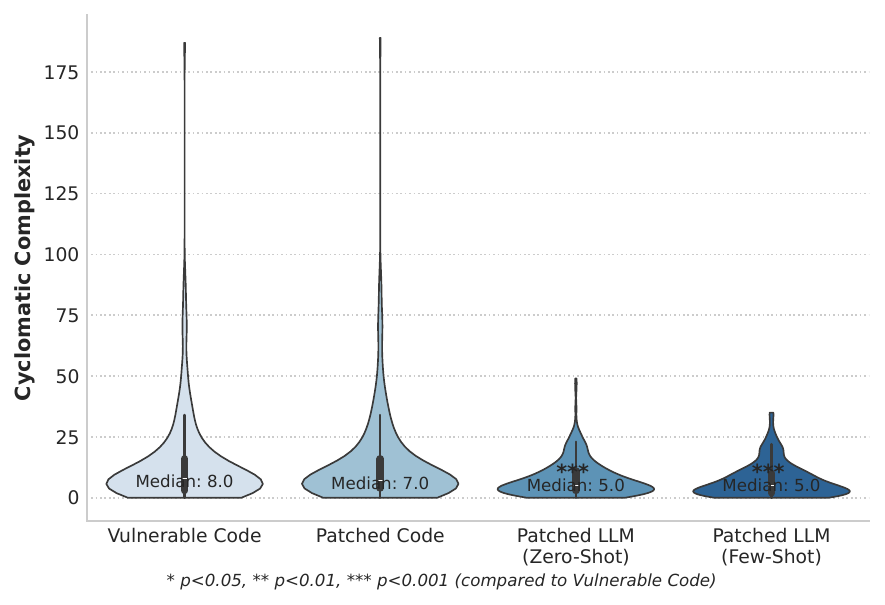}}
\caption{Cyclomatic Complexity distribution across code types. * p<0.05, ** p<0.01, *** p<0.001 (vs. Vulnerable Code). LLM patches show significantly lower complexity, suggesting simplified solutions that may miss critical vulnerability aspects.}
\label{figure:RQ4-complexity}
\end{figure}

Figure \ref{figure:RQ4-complexity} presents an analysis of the cyclomatic complexity for vulnerable code, patched code, and LLM-generated patches in both zero-shot and few-shot scenarios. The results reveal that the average complexity of LLM-generated patches is generally lower than that of both the vulnerable and patched code. Specifically, the median cyclomatic complexity of vulnerable code (8) and patched code (7) is significantly higher than LLM-generated patches in both zero-shot and few-shot scenarios (both with median complexity of 5). This highlights a limitation in LLM-generated patches, as they often fail to capture the nuanced complexity required to provide comprehensive fixes, potentially resulting in partial or insufficient repairs for complex vulnerabilities.

As mentioned in Section \ref{subsec:repair}, our weighted aggregation system assigned 30\% of the total scoring weight to cyclomatic complexity, reflecting the importance of maintaining manageable complexity in patch generation. This weighting is flexible and can be adjusted based on the specific characteristics of the vulnerabilities being addressed. For instance, when addressing intricate vulnerabilities that require more complex logical corrections, the weight assigned to cyclomatic complexity could be increased to prioritize comprehensive fixes over simplicity.

To address the complexity deficit in LLM-generated patches, models with larger context windows are essential, as they can process entire codebases simultaneously, better preserving cross-function dependencies and control flow structures that smaller-context models often simplify or overlook.

\begin{figure}[b]
\centerline{\includegraphics[width=0.9\columnwidth]{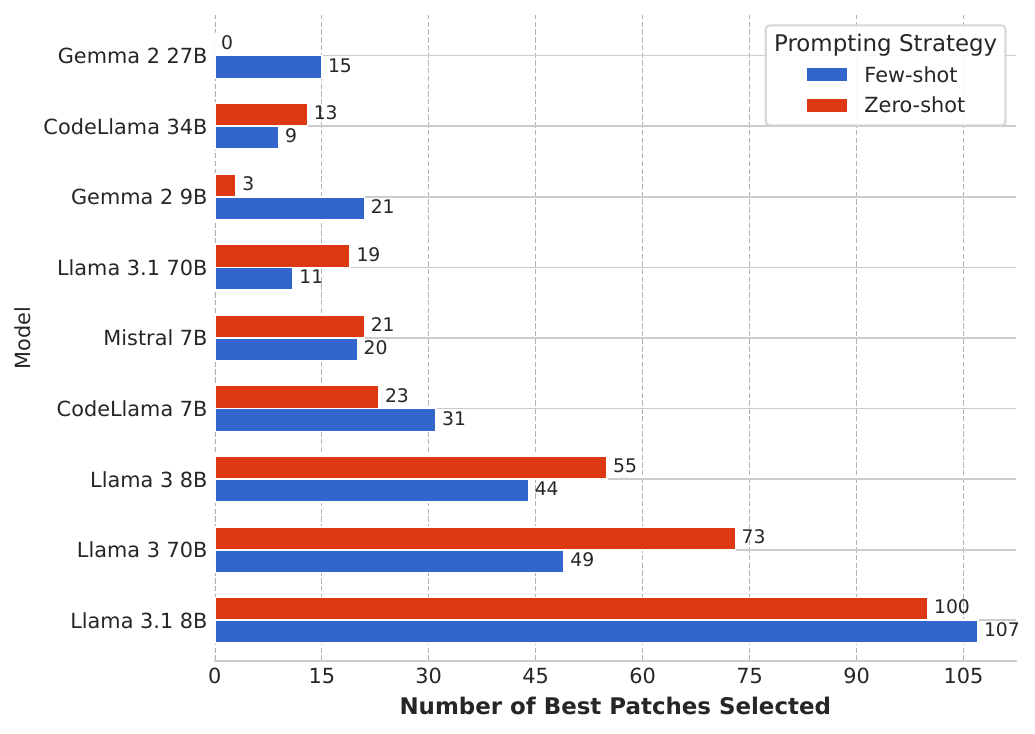}}
\caption{Best Models for SVR Tasks}
\label{figure:SVR-bestmodel}
\end{figure} 

As shown in Figure \ref{figure:SVR-bestmodel}, models like \texttt{llama3.1 8b}, \texttt{llama 3 70b}, and \texttt{llama3 8b} consistently provided high-quality patches. Note that, while our weighted aggregation strategy improved patch quality overall, complex cases remain challenging. Figure~\ref{fig:cve2016_patch} illustrates a permission vulnerability (CWE-284) where Llama3-8B produced a structurally consistent patch but missed essential security fixes: replacing unsafe \texttt{ACL} operations and implementing proper inode locking. Even with moderate cyclomatic complexity (7), the function's multiple control branches and error paths complicated reasoning about synchronization semantics. This highlights a common LLM limitation: generating patches that appear correct but lack critical security logic for complex vulnerabilities. Potential remedies include security-focused fine-tuning and providing explicit vulnerability patterns as additional context in prompts.

\begin{figure}[t]
\centering
\begin{lstlisting}[style=codestyle]
// File path: fs/nfsd/nfs2acl.c (similar changes in nfs3acl.c and nfs4acl.c)
static __be32 nfsacld_proc_setacl(struct svc_rqst * rqstp, ...)
{
    // ...
    inode = d_inode(fh->fh_dentry);
@DEL@-   if (!IS_POSIXACL(inode) || !inode->i_op->set_acl) {@ENDDEL@
@DEL@-       error = -EOPNOTSUPP;@ENDDEL@
@DEL@-       goto out_errno;@ENDDEL@
@DEL@-   }@ENDDEL@

    error = fh_want_write(fh);
    if (error)
        goto out_errno;

@ADD@+   fh_lock(fh);@ENDADD@

@DEL@-   error = inode->i_op->set_acl(inode, argp->acl_access, ACL_TYPE_ACCESS);@ENDDEL@
@ADD@+   error = set_posix_acl(inode, ACL_TYPE_ACCESS, argp->acl_access);@ENDADD@
    if (error)
@DEL@-       goto out_drop_write;@ENDDEL@
@DEL@-   error = inode->i_op->set_acl(inode, argp->acl_default,@ENDDEL@
@DEL@-                    ACL_TYPE_DEFAULT);@ENDDEL@
@ADD@+       goto out_drop_lock;@ENDADD@
@ADD@+   error = set_posix_acl(inode, ACL_TYPE_DEFAULT, argp->acl_default);@ENDADD@
    if (error)
@DEL@-       goto out_drop_write;@ENDDEL@
@ADD@+       goto out_drop_lock;@ENDADD@

@ADD@+   fh_unlock(fh);@ENDADD@

    // Error handling
    return nfserr;
@DEL@-out_drop_write:@ENDDEL@
@ADD@+out_drop_lock:@ENDADD@
@ADD@+   fh_unlock(fh);@ENDADD@
    fh_drop_write(fh);
    // ...
}
\end{lstlisting}
\tiny
\caption{Patch for \textbf{CVE-2016-1237} (\textbf{CWE-284}): Permission bypass vulnerability in NFS ACL handling with high code complexity (143 lines in LLM patch). The fix replaces direct ACL operations with permission-checked alternatives and adds proper inode locking. Highlights the challenge of generating semantically correct patches for complex vulnerabilities. %permission-related vulnerabilities that span multiple files with similar patterns.
}
\label{fig:cve2016_patch}
\end{figure}

\section{Limitations and Directions}
\label{section:discussion}

% Our study has key limitations of ensemble-based LLM approaches: (1) patch evaluation is based on similarity metrics, which may not accurately reflect functional correctness; and (2) component selection prioritizes model diversity without accounting for individual performance. \mv{these do not seem limitations of the ensemble-based LLM approaches, but of our paper}

% Regarding threats to validity, the inherent non-determinism of LLM outputs introduces variability in results. Although the dataset was curated from CVE-linked commits with measures taken to exclude non-vulnerability changes, potential labeling inconsistencies may still affect evaluation accuracy. To mitigate data leakage, we sourced code from different versions of the Linux kernel; however, this cannot fully eliminate the risk.

% Our study has several limitations. Patch evaluation relies primarily on similarity metrics, which may not accurately reflect the functional correctness of generated patches. Additionally, our ensemble approach emphasizes model diversity in component selection without accounting for individual model performance, potentially limiting its overall effectiveness. 
% The inherent non-determinism of LLM outputs introduces variability in the results, and although the dataset was curated from CVE-linked commits and filtered to exclude non-vulnerability-related changes, labeling inconsistencies may still affect evaluation accuracy. Furthermore, while we sourced code from different versions of the Linux kernel to mitigate the risk of data leakage, this does not entirely eliminate the possibility.

Our study has several limitations. Evaluation based on similarity metrics may not fully capture the functional correctness of patches. The ensemble approach prioritizes model diversity over individual performance, which may limit effectiveness. Output variability due to LLM non-determinism and possible labeling inconsistencies in the curated CVE dataset could affect accuracy. While we used different Linux kernel versions to reduce data leakage, the risk cannot be entirely ruled out.

Future research should explore: (1) adaptive weighting mechanisms that dynamically adjust model contributions based on historical performance; (2) integrating static analysis feedback to leverage both symbolic and neural methods; and (3) uncertainty-aware methods that express confidence levels rather than binary decisions.

\section{Related Work}
\label{section:related-works}

Recent studies have examined LLMs' capabilities for vulnerability detection and repair, with researchers like Ullah et al. \cite{ullah2024llms}, Khare et al. \cite{khare2023understanding}, Sun et al. \cite{sun2024llm4vuln}, and Zibaeirad et al. \cite{zibaeirad2025reasoning} evaluating individual models on specific vulnerability patterns. 
Despite promising results, these studies consistently highlight the challenge of model inconsistency, where even advanced LLMs struggle with complex vulnerabilities and produce conflicting results across similar tasks. While benchmarking efforts from Liu et al. \cite{liu2024vuldetectbench}, Gao et al. \cite{gao2023far}, and Ding et al. \cite{ding2024vulnerability} have expanded evaluation datasets, they focus primarily on single-model performance rather than addressing this fundamental inconsistency problem. 

Traditional approaches to vulnerability detection have evolved from static analysis tools \cite{pearce2022asleep} to dynamic testing \cite{gan2018collafl,bohme2016coverage} and machine learning techniques \cite{chu2024graph,wen2023vulnerability,wang2020combining,zibaeirad2024comprehensive,zhou2019devign,rashid2024quantifying,babaey2024gensqli,babaey2025detecting,babaey2025genxss}, with recent hybrid methods \cite{li2024llm,liu2024pre,wang2024combining} attempting to combine their strengths. DVDR-LLM distinguishes itself from these works by introducing a novel ensemble-based framework that explicitly addresses model inconsistency through diversity, offering a comprehensive study that systematically aggregates the strengths of multiple LLM families across both detection and repair tasks, complete with a weighted repair evaluation system that considers code quality beyond syntactic similarity, and a unique evaluation across three levels of code abstraction to provide practical insights into how ensemble techniques scale with real-world complexity—capabilities not addressed by previous research focused on individual model performance.

\section{Conclusion}
\label{section:conclusion}
This study highlights the limitations of individual LLMs in handling complex vulnerabilities and presents DVDR-LLM as an ensemble-based approach to improve detection and repair. By aggregating diverse model outputs, DVDR-LLM reduces reliance on single-model predictions and enhances the robustness of software vulnerability detection and repair. While it shows promise in improving accuracy and repair quality, challenges remain. Future research should focus on multi-agent LLM systems, reinforcement learning, and integrating static and dynamic analysis tools to address these gaps and advance the applicability of LLMs in software security tasks.

%% Acknowledgments (optional)
% \begin{acks}
% This work was supported by...
% \end{acks}

%% Bibliography - ACM style
\bibliographystyle{ACM-Reference-Format}
\bibliography{ref}

%% Balance columns on last page
\balance

\end{document}